\documentclass[conference]{IEEEtran}
\IEEEoverridecommandlockouts
\usepackage{amsthm} 

\usepackage{algorithm}
\usepackage{algorithmic}
\usepackage{amsmath,amsfonts}
\allowdisplaybreaks[4]
\usepackage{algorithmic}
\usepackage{algorithm}
\usepackage{array}
\usepackage[caption=false,font=normalsize,labelfont=sf,textfont=sf]{subfig}
\usepackage{textcomp}
\usepackage{stfloats}
\usepackage{url}
\usepackage{verbatim}
\usepackage{graphicx}
\usepackage{cite}
\usepackage{amssymb}
\usepackage{nomencl}
\makenomenclature
\usepackage{etoolbox}
\usepackage{url}

\def\BibTeX{{\rm B\kern-.05em{\sc i\kern-.025em b}\kern-.08em
    T\kern-.1667em\lower.7ex\hbox{E}\kern-.125emX}}

\makeatletter
\newcommand{\linebreakand}{%
  \end{@IEEEauthorhalign}
  \hfill\mbox{}\par
  \mbox{}\hfill\begin{@IEEEauthorhalign}
}
\makeatother

\begin{document}
\title{Energy-grade double pricing mechanism for a combined heat and power system using the asynchronous dispatch method\\
\thanks{This work was supported in part by the National Key R\&D Program of China under Grant 2020YFB0906000 and 2020YFB0906005. Corresponding author: Ye Guo, e-mail: guo-ye@sz.tsinghua.edu.cn.}
\thanks{This work has been submitted to the 2022 IEEE 6th Conference on Energy internet and Energy System Integration for possible publication. Copyright may be transferred without notice, after which this version may no longer be accessible.}
}

\author{\IEEEauthorblockN{1\textsuperscript{st} Xinyi Yi}
\IEEEauthorblockA{\textit{Tsinghua-Berkeley Shenzhen Institute} \\
\textit{Tsinghua University}\\
Guangdong 518055, China \\
yi-xy20@mails.tsinghua.edu.cn}
\and
\IEEEauthorblockN{2\textsuperscript{nd} Ye Guo$^*$}
\IEEEauthorblockA{\textit{Tsinghua-Berkeley Shenzhen Institute} \\
\textit{Tsinghua University}\\
Guangdong 518055, China \\
guo-ye@sz.tsinghua.edu.cn}
\and
\IEEEauthorblockN{3\textsuperscript{rd} Hongbin Sun}
\IEEEauthorblockA{\textit{Department of Electrical Engineering} \\
\textit{Tsinghua University}\\
Beijing 100000, China \\
shb@tsinghua.edu.cn}
\linebreakand 
\IEEEauthorblockN{4\textsuperscript{th} Qiuwei Wu}
\IEEEauthorblockA{\textit{Tsinghua-Berkeley Shenzhen Institute} \\
\textit{Tsinghua University}\\
Guangdong 518055, China \\
qiuwu@sz.tsinghua.edu.cn}
\and
\IEEEauthorblockN{5\textsuperscript{th} Li Xiao}
\IEEEauthorblockA{\textit{Tsinghua-Berkeley Shenzhen Institute} \\
\textit{Tsinghua University}\\
Guangdong 518055, China \\
xiaoli@sz.tsinghua.edu.cn}
}
\maketitle

\begin{abstract}
The problem of heat and electricity pricing in
combined heat and power (CHP) systems regarding the time scales of electricity and heat, as well as thermal energy quality, is studied. Based on the asynchronous coordinated dispatch of the CHP system, an energy-grade double pricing mechanism is proposed. Under the pricing mechanism, the resulting merchandise surplus of the heat system operator (HSO) at each heat dispatch interval can be decomposed into interpretable parts and its revenue adequacy can be guaranteed for all heat dispatch intervals. And the electric power system operator (ESO)'s resulting merchandise surplus is composed of non-negative components at each electricity dispatch interval, also ensuring its revenue adequacy. In addition, the effects of different time scales and co-generation
are analyzed in different kinds of CHP units’ pricing.
\end{abstract}


\section{Introduction}
To achieve the objectives of energy saving and carbon emission reduction, the CHP system has been deployed all around the world due to its high fuel efficiency \cite{2007Efficiency} and remarkable complementary \cite{2017Coordinated}. However, the two systems with different properties and time scales are interdependent on CHP units' co-generation. The generation costs of heat and electricity are also coupled for CHP units. Therefore, the pricing of electricity and heat for CHP systems needs to be further studied.

Different from the widely adopted LMP in the electricity market, the pricing of heat is an open question. Heat pricing can be divided into two categories: cost-plus pricing which is often adopted in regulated heat markets like Russia, and the marginal cost pricing method which is commonly used in deregulated heat markets like the Netherlands. The cost-plus pricing model aims to recover the sum of heat generation companies' cost and provide reasonable profits. Marginal cost pricing is based on the cost of an additional unit of heat in the heat system (HS). It encourages heat generation companies to compete with each other and the company may gain less profit than intended. Since marginal cost pricing reflects the scarcity of resources, it becomes the mainstream of the research on heat pricing. Paper \cite{2019The} proposes heat marginal energy prices in a competitive wholesale heat market. Paper \cite{2019A} uses heat energy's levelized cost for pricing, considering the average fixed cost of the energy plant and marginal generation cost.

The pricing of heat and electricity in the CHP system is also extensively studied. Paper \cite{2019Generalized} extends LMP into heat pricing based on the synchronous coordinated dispatch (CD) for the CHP system. Heat and electricity are priced as the shadow prices related to nodal heat balance and bus energy balance. Papers \cite{9615373,8931731} study the formulation of heat and electricity prices based on CD for the CHP system considering integrated demand response (IDR).

However, the above research is based on the synchronous CD for the CHP system. But electric power system (EPS) is adjusted in minutes to achieve real-time power balance, whereas HS has higher inertia and is usually adjusted in hours. In \cite{2017Coordinated} the equipment of EPS follows the electricity dispatch time scale, and the equipment of HS follows the heat dispatch time scale. Paper \cite{9220970} dispatches the variables of EPS in minutes and the variables of HS in hours. Compared to \cite{2017Coordinated}, \cite{9220970} reduces the total generation cost by fully using the flexibility of CHP units. 

Another limitation of the current study which extends LMP into heat pricing directly in the CHP system is that it is under the assumption of heat energy's homogeneity. But heat energy has different grades according to heat carriers' temperatures, and would influence both the heat generators' efficiency \cite{6041454} and heat consumers' comfort \cite{6477323}. An energy-grade double pricing mechanism in the standalone heat market is presented in our previous work \cite{2015arXiv150203167I} to settle both heat energy and grades. But it does not consider the CHP system with co-generation and different time scales. Thus, the pricing mechanism in the CHP system considering the two energy carriers' characteristics is studied in this paper.

The major contributions of this paper are as follows:

1) An electricity and heat joint market clearing model adopting hybrid time scales is established. A pricing mechanism for heat and electricity based on the model is proposed.

2) Revenue adequacy for operators of both the EPS and HS is established in the proposed pricing mechanism. 

3) The components of heat and electricity prices are studied in different kinds of CHP units, which are related to CHP units' feasible regions and reflect the different time scales.

\section{Problem modeling}
In this section, we present an asynchronous coordinated dispatch model for the CHP system.

\subsection{Different time scales of the EPS and HS}
\begin{figure}[htbp]
	\vspace{-0.3cm}
	\hspace{-0.2cm}
	\includegraphics[width=3.3in]{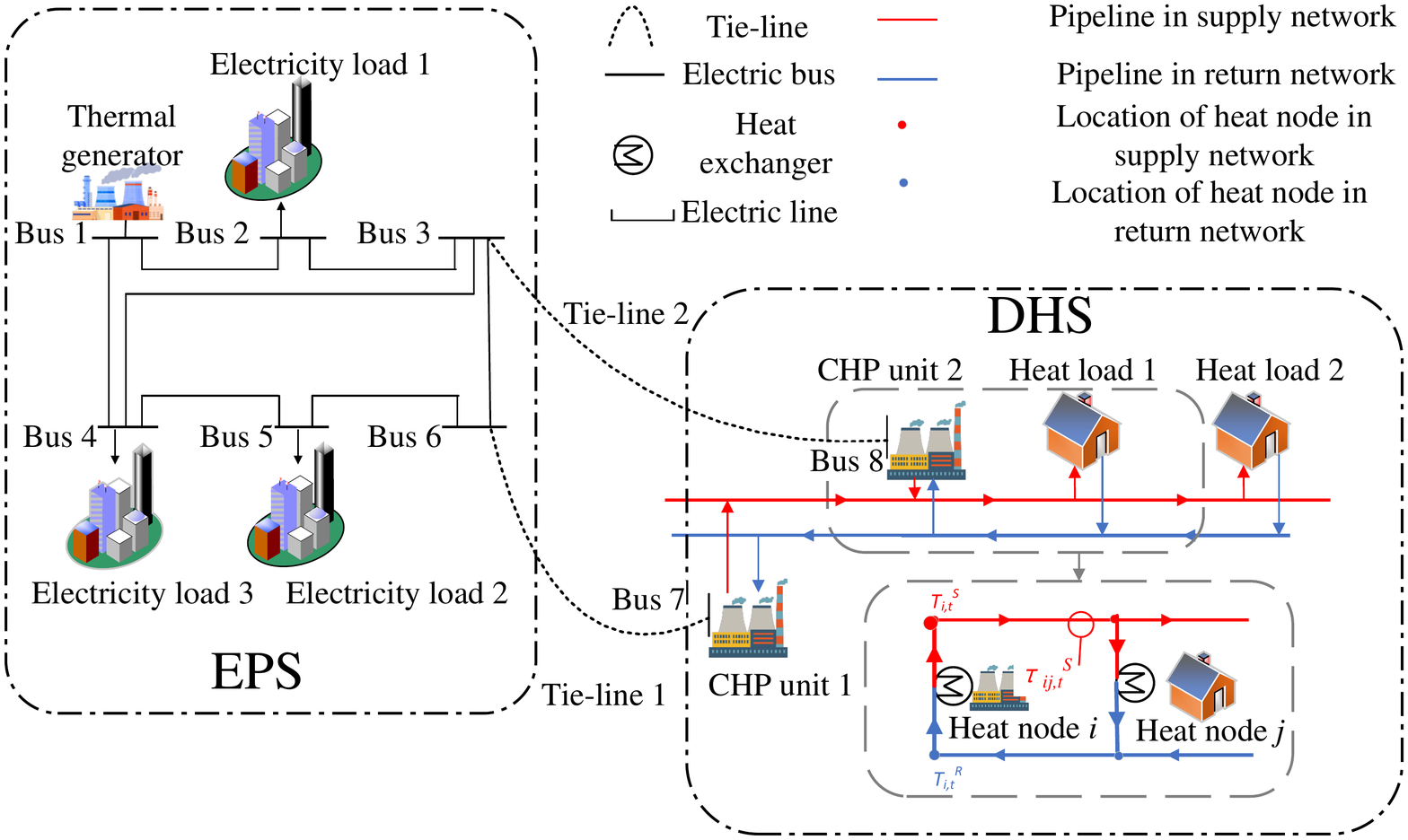}
	\vspace{-1.2cm}
	\caption{The general structure of a CHP system.}
	\label{fig 1}
\end{figure}

The EPS and HS are coupled by CHP units as shown in Fig. \ref{fig 1}. They have very different time scales.

 \vspace{-0.3cm}

    \begin{figure}[htbp]
	\vspace{-0.7cm}
	\hspace{-0.2cm}
	\includegraphics[width=3.6in]{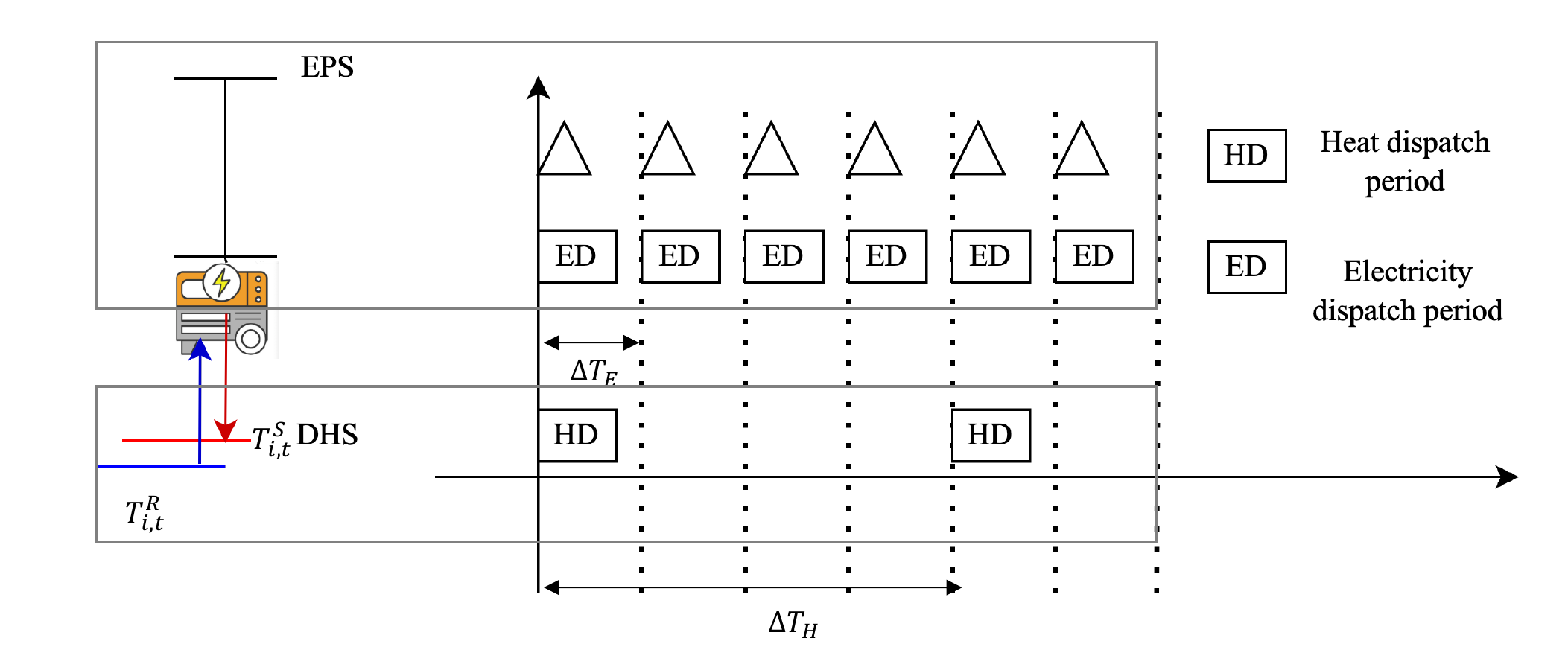}
	\vspace{-0.8cm}
	\caption{Different time scales.}
	\label{fig 2}
    \end{figure}
	\vspace{-0.4cm}

We use the asynchronous dispatch model in \cite{9220970} which dispatches variables of EPS and HS in minutes and hours, respectively as shown in Fig. \ref{fig 2}, where $\Delta T_E$ and $\Delta T_H$ represent the dispatch interval of EPS and HS. For simplicity of analysis later, we assume that the $\Delta T_H$ divides $\Delta T_E$ exactly. The variables' time scales are shown in TABLE \ref{table1}. We can see that hybrid time scales are adopted for the CHP units. 
\begin{table}[ht]
\vspace{-0.0cm}
\hspace{-6.6cm}
	\footnotesize
	\centering
	\setlength{\abovecaptionskip}{-0.5pt}
	\setlength{\belowcaptionskip}{0pt}
	\caption{Time scales and operators of variables}\label{table1}
	\begin{tabular}{p{20pt}p{90pt}p{40pt}p{20pt}p{25pt}}
		\hline
		Variables&Definition&Operator&Time scale\\
		\hline
		$\boldsymbol{\delta_t}$&Bus angles&EPS&$\Delta T_E$\\
		$\boldsymbol{G_{p,t}^E}$&Pure electricity generator electricity output&EPS&$\Delta T_E$\\
		$\boldsymbol{G_{p,t}^C}$&CHP unit electricity output&EPS&$\Delta T_E$\\
		$\boldsymbol{G_{h,t}^C}$&CHP unit heat output&DHS&$\Delta T_H$\\
		$\boldsymbol{G_{h,t}^H}$&Heat boiler heat output&DHS&$\Delta T_H$\\
		$\boldsymbol{T_t}$&Locational temperatures&DHS&$\Delta T_H$\\
		\hline
	\end{tabular}
	\end{table}

\subsection{Operation model}
In the section, we introduce the heat network which is dispatched in $\Delta T_H$, the electricity network which is dispatched in $\Delta T_E$, and energy sources dispatched in hybrid time scales.
\subsubsection{Heat network formulation}
A heat system includes heat exchangers, a supply network, and a return network. We use $ \tau$ and $T$ to distinguish the temperatures of the pipeline's outlet and exchangers' connecting locations in the following parts.  

The node injects/consumes heat power through the heat exchanger between the supply and return networks. The source/load nodal heat balance is presented as the exchanger's supply/return-side temperature mixing process \cite{2019Generalized}:
\begin{equation}\label{nb}
	\begin{aligned}
		G_{hi,t}-D_{hi,t}=&c[(M_{i,t}+\sum_{ji\in	\Psi_{S/R}(i)}m_{ji,t})T_{i,t}^{S/R}-M_{i,t}	T_{i,t}^{R/S}\\
		&-\sum_{ji\in\Psi_{S/R}(i)}m_{ji,t}\tau_{ji,t}^{S/R}],\forall i\in\Psi_{HS/HL}.
	\end{aligned}
\end{equation}
Each heat node participates in temperature mixing processes on its two sides. Temperature mixing processes in source node $i$'s return side and load node $i$'s supply side are:
\begin{equation}
\begin{split}
		0=(\sum_{ji\in \Psi_{R/S}(i)}m_{ji,t})T_{i,t}^{R/S}-\sum_{j\in \Psi_{R/S}(i)}m_{ji,t}\tau_{ji,t}^{R/S}, \\
		\forall i\in\Psi_{HS/HL}.\label{tm}
		\end{split}
\end{equation}
In a pipeline, temperature changes at its inlet node are transferred to the outlet node slowly, which takes about the transport time of mass flow through the pipeline. This is why HS has a larger time scale than EPS. $\tau_{ji,t}^{S/R}$ is calculated as:
\begin{equation}
\begin{split}
	\tau_{ji,t}^{S/R}=&((1-\psi_{ji}^{S/R})T_{j,t}^{S/R}+\psi_{ji}^{S/R}T_{j,t-1}^{S/R}-T_{ak,t}^{S/R})\\&*(1-v_{rk}^{S/R}*L_{rk}^{S/R}/cm_{rk,t}^{S/R})+T_{ak,t}^{S/R}. 
	\end{split}\label{caltem}
\end{equation}
$\psi_{ji}^{S/R}$ is the transfer time from node $j$ to node $i$ in supply/return network respectively, for its detailed calculation, please refer to \cite{2015arXiv150203167I}. $T_{i,t}^{S/R}$ represent locational temperature of node $i$ in supply/return network at heat dispatch period $t$ respectively. $M_{i,t}$ is the mass flow through the nodal heat exchanger at node $i$ at period $t$, and $m_{ji,t}$ is the mass flow through pipeline $ji$. $c$ is the specific heat capacity of the water. $G_{hi,t}$ and $D_{hi,t}$ represent heat node $i$'s heat production and load power respectively.  $\Psi_{S/R}(i)$ represent the set of pipelines injecting into the supply/return locations of node $i$ respectively. $\Psi_{HL/HS}$ are the set of heat load/source nodes respectively. For the detailed description of the heat exchange and transfer process, please refer to \cite{2015arXiv150203167I}.

\subsubsection{Electric network formulation}
Bus $i$'s power balance equation is shown in (\ref{4}). $\Psi_{EB}$ is the set of buses in EPS.
\begin{equation}\label{4}
G_{pi,t}-D_{pi,t}-\sum_{r\in\Psi_{EB}}\frac{1}{x_{ir}}(\delta_{i,t}-\delta_{r,t})=0,
\end{equation}
where $x_{ir}$ presents the reactance of electric power line $ir$.

\subsubsection{Energy sources}
Feasible regions of different kinds of CHP units can be described by polytopes \cite{2017Coordinated} as:
\begin{equation}
    \boldsymbol{O_i}G_{pi,t}+\boldsymbol{K_i}G_{hi,r}\le \boldsymbol{V_i}, t\in N_E, r=t\frac{N_H}{N_E},  \label{gen}
\end{equation}
where $N_H$ and $N_E$ are the numbers of heat and electricity dispatch intervals respectively. There are two kinds of CHP units-extraction condensing and back-pressure CHP units. Their feasible regions are shown in Fig. 3. The electric and heat power outputs of back-pressure units have a linear relationship together with the capacity constraints. We present the linear relationship as a pair of inequations to fit in (\ref{gen}).
      \begin{subequations}
      \begin{align}
        &-O_{i,1}G_{pi,t}-K_{i,1}G_{hi,r}\le -V_{i,1},\\
        &O_{i,1}G_{pi,t}+K_{i,1}G_{hi,r}\le V_{i,1}, t\in N_E, r=\frac{N_H}{N_E},\\
        &\underline {G_{hi}}\le G_{hi,t}, t\in N_H,\\
        &G_{hi,t}\le \overline {G_{hi}}, t\in N_H.
        \end{align}
    \end{subequations}

	\begin{figure}[htbp]
	\vspace{-0.7cm}
	\hspace{0.4cm}
	\includegraphics[width=3.0in]{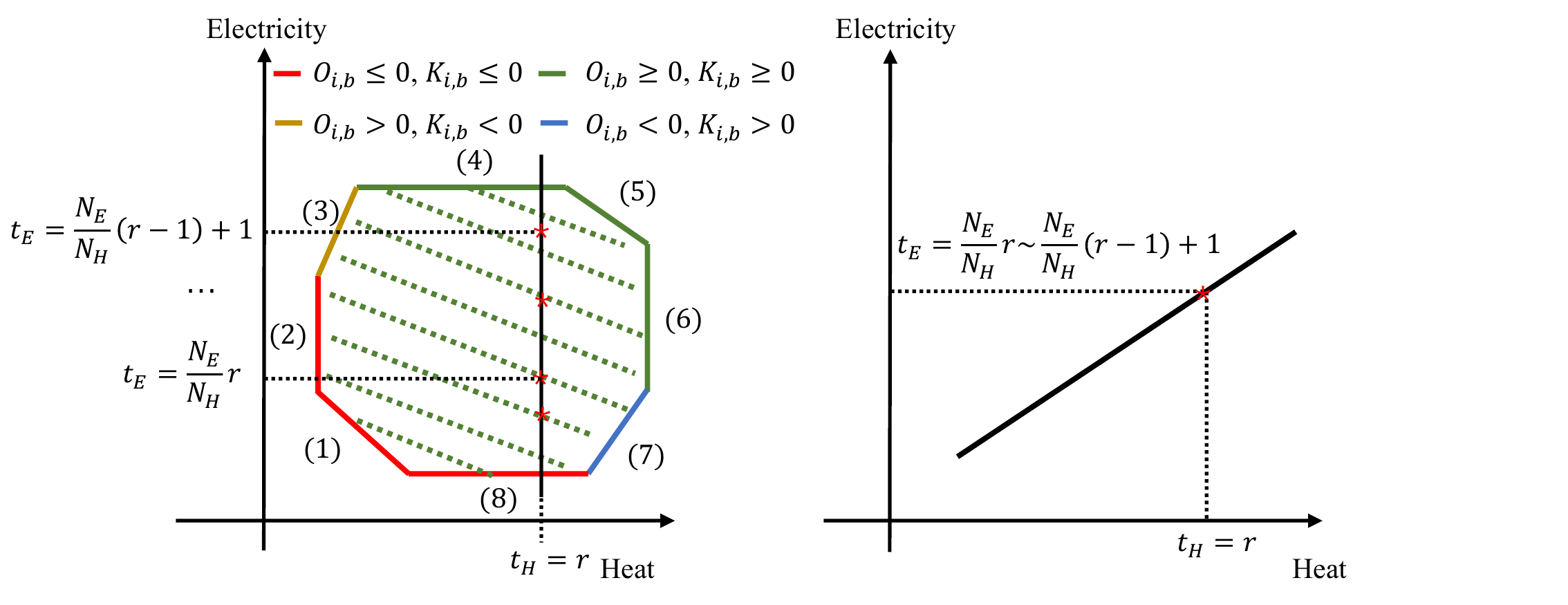}
	\vspace{-0.3cm}
	\caption{Feasible regions of two kinds of CHP units.}
	\label{fig3}
    \end{figure}
    \vspace{-0.2cm}
Specially, for pure electricity-output and heat-output units whose operation constraints are $\underline G_{pi}\le G_{pi,t}\le \overline G_{pi}$ and $\underline G_{hi}\le G_{hi,t}\le \overline G_{hi}$ are special examples of (5).

\subsection{Asynchronous dispatch model for CHP systems}
Using the strategy of variable temperatures and fixed mass flow in the heat network \cite{2017Coordinated}\cite{2019Generalized}\cite{9220970}, the optimal asynchronous dispatch model of CHP systems is formulated as:
\begin{equation}\label{object}
\begin{aligned}
	&\min_{\boldsymbol{G_{h,t},G_{p,t},T_{t},\delta_{t}}} 
	f=\sum_{t=1}^{T_E}(\sum_{i\in \Psi_{CHP}} \mathcal{C}_C(G_{pi,t},G_{hi,r})\\
	&+\sum_{i\in \Psi_{TG}} \mathcal{C}_{E}(G_{pi,t}))*\Delta T_E+\sum_{t=1}^{T_H}(\sum_{i\in \Psi_{HB}}\mathcal C_{H}(G_{hi,t})\\
	&+\sum_{i\in \Psi_{CHP}}\mathcal C_{C}(G_{hi,t}))*\Delta T_H, r=t\frac{N_H}{N_E}.\\
	\end{aligned}
	\end{equation}
$s.t.$
\begin{subequations}\label{cons}
    \begin{align}
     &\boldsymbol \lambda_{h,t}:\boldsymbol {C_{1}T_{t}+C_{2}T_{t-1}+R_{t}}=\boldsymbol {H_t}, t\in N_H,\label{nbm}\\
	 &\boldsymbol \mu_{t}: \boldsymbol {T_{t}\le T_{sa}}, t\in N_H,\label{ts}\\
	 &\boldsymbol \beta_{t}: \boldsymbol {T_{t}\ge T_{Q}}, t\in N_H,\label{tq}\\
	 &\boldsymbol \lambda _{p,t}:\boldsymbol{B\delta_{t}}=\boldsymbol{G_{p,t}-D_{p,t}}, t\in N_E,\label{eb}\\
	 &\boldsymbol \sigma_t:\boldsymbol{F\delta_{t}\le \overline L}, t\in N_E,\label{lp}\\
	&\boldsymbol{\gamma_{t}}:\boldsymbol{OG_{p,t}+KG_{h,r}\le V}, t\in N_E, r=t\frac{N_H}{N_E}.\label{fr}
	\end{align}
\end{subequations}

\subsubsection{Objective function}
The objective function (\ref{object}) is to minimize the total cost of generators in all time intervals. The generators include pure electricity/heat-output unit $i$ whose cost function is $\mathcal{C}_{Ei,t}$/$\mathcal{C}_{Hi,t}$, and CHP unit $i$ whose cost function is $\mathcal{C}_{Ci,t}$. The cost functions are expressed using a quadratic function of electricity/heat power production $G_{pi,t}/G_{hi,t}$:
\begin{subequations}
    \begin{align}
    \mathcal{C}_{Ei,t}=&\begin{matrix} 
    \underbrace{\eta_{i0}^E+\eta_{i1}^EG_{pi,t}+\eta_{i2}^EG_{pi,t}^2}\\
    \mathcal{C}_{E}(G_{pi,t})
    \end{matrix},\\
    \mathcal{C}_{Hi,t}=&\begin{matrix} 
    \underbrace{\eta_{i0}^H+\eta_{i1}^HG_{hi,t}+\eta_{i2}^HG_{hi,t}^2}\\
    \mathcal{C}_{H}(G_{hi,t})
    \end{matrix},\\
    \mathcal{C}_{Ci,t}=&\begin{matrix} 
    \underbrace{\eta_{i0}^C+\eta_{i1}^CG_{hi,t}+\eta_{i2}^CG_{hi,t}^2}\\
    \mathcal{C}_{C}(G_{hi,t})
    \end{matrix}\\
    &+\begin{matrix} 
    \underbrace{\eta_{i3}^CG_{pi,t}
    +\eta_{i4}^CG_{pi,t}^2}\\\notag
    \mathcal{C}_{C}(G_{hi,t})
    \end{matrix}+\begin{matrix} 
    \underbrace{\eta_{i5}^CG_{pi,t}G_{hi,t}}\\
    \mathcal{C}_{C}(G_{hi,t},G_{pi,t})\end{matrix}.
    \end{align}
\end{subequations}

\subsubsection{Constraints}
\begin{itemize}
    \item Nodal balancing \& temperature mixing constraints (\ref{nbm}): It is the matrix form of (\ref{nb}-\ref{tm}). 
    \begin{equation}
	\begin{aligned}
	    \boldsymbol C_{1}&=\boldsymbol {A_{a1}G_{c}A_{a1}^{T}-A_{a2}G_{c}DaA_{a1}^{T}},\\
		\boldsymbol C_{2}&=\boldsymbol {-A_{a2}G_{c}DbA_{a1}^{T}},\\
		\boldsymbol{D}&=\boldsymbol I-\boldsymbol v \boldsymbol{L}/c\boldsymbol{m},\boldsymbol G_{c}=c\boldsymbol{m},\\
		\boldsymbol a&=\boldsymbol I-\rho \boldsymbol{S} \boldsymbol{L} /(\boldsymbol{m} \Delta T),\boldsymbol b= \rho \boldsymbol{S} \boldsymbol{L} /(\boldsymbol{m} \Delta T),\\
		\boldsymbol R&=\boldsymbol {A_{a2} G_{c} (D-I) A_{a1}^T T_{a,t}}
	\end{aligned}
\end{equation}
where $\boldsymbol{S}$, $\boldsymbol{L}$ and $\boldsymbol{m}$ are diagonal matrices of pipelines' cross-section area, length and mass flow rates. $\boldsymbol I$ and $\boldsymbol 1$ are diagonal matrix and vector whose element is 1. For its detailed derivation, please refer to \cite{2015arXiv150203167I}.
    \item Pipeline security constraints on temperature (\ref{ts}):
    To ensure the heat network working conditions, the temperature should not exceed $\boldsymbol{T_{sa}}$.
    \item Temperature requirements of heat sources and consumers (\ref{tq}): Heat consumers may have temperature requirements to ensure comfort. Heat sources may also propose temperature requirements to enhance production efficiency.
    \item Electric power balance (\ref{eb}), where $\boldsymbol{G_{p,t}}$ and $\boldsymbol{D_{p,t}}$ represent the electric output and load power respectively.
    \item Transmission capacity constraints (\ref{lp}).
    \item Energy sources constraints (\ref{fr}).
\end{itemize}

\section{Energy-grade double pricing mechanism}
In our previous work\cite{2015arXiv150203167I}, we prove that a direct extension of LMP into heat pricing leads to the HS operator's negative merchandise surplus because that heat has the issue of grade. Thus we extend the energy-grade double pricing mechanism proposed in \cite{2015arXiv150203167I} which settles both the heat energy and grade into the CHP system considering different time scales. 
\subsection{Pricing and settlement}
The Lagrangian of the asynchronous CD (\ref{object}-\ref{cons}) is:
\begin{equation}
	\begin{aligned}
		L&=f+\sum_{t=1}^{N_H}[\boldsymbol {\lambda_{h,t}^T(C_{1}T_{t}+C_{2}T_{t-1}+R_{t}-H_{t})}\\
		&+\boldsymbol{\mu_{t}^T(T_{t}-T_{sa})+\beta_{t}^T(T_{Q}-T_{t})}]+\sum_{t=1}^{N_E}[\boldsymbol {\lambda_{p,t}^T(B\delta_t}\\
		&\boldsymbol{-G_{p,t}+D_{p,t})+\sigma_t^T(F\delta_t-\overline{L})}+\boldsymbol{\gamma_{t}^T(OG_{p,t}}\\
		&\boldsymbol{+KG_{h,r}-V)}],r=\frac{N_H}{N_E}.\\
	\end{aligned}
\end{equation}

If heat node $i$'s heat balance equation corresponds to $k^{th}$ row of (\ref{nbm}). The heat energy marginal price of node $i$ at $t$ is:
\begin{equation}\label{he}
\begin{aligned}
	\varrho_{h,t}^i&= \frac{\partial f^*}{\Delta T_H \partial D_{hi,t}}=\frac{\partial L}{\Delta T_H \partial D_{hi,t}}=-\frac{\partial L}{\Delta T_H \partial H_{hk,t}}\\
	&=\frac {\lambda_{hk,t}^*}{\Delta T_H}, t\in N_H.
	\end{aligned}
\end{equation}

Heat node $i$'s temperature requirements are settled at heat grade marginal prices at $t$:
\begin{subequations}\label{hg}
\begin{align}
	{\varrho_{g,t}^{Si}}&=\frac{\partial f^*}{\Delta T_H\partial T_{Qi,t}^S}=\frac{\partial L}{\Delta T_H\partial  T_{Qi,t}^S}=\frac{\beta_{i,t}^{S*}}{\Delta T_H}, t\in N_H.\\
	{\varrho_{g,t}^{Ri}}&=\frac{\partial f^*}{\Delta T_H\partial T_{Qi,t}^R}=\frac{\partial L}{\Delta T_H\partial  T_{Qi,t}^R}=\frac{\beta_{i,t}^{R*}}{\Delta T_H}, t\in N_H.
	\end{align}
\end{subequations}

Thus the payment of load at node $i$ in the heat market is:
\begin{equation}
	\begin{split}
		\pi_{i}^{HL}&=\Delta T_H\sum_{t=1}^{N_H}\pi_{i,t}^{HL}=\Delta T_H\sum_{t=1}^{N_H}[(\varrho_{h,t}^{i*}*D_{hi,t})+(\varrho_{g,t}^{Si*}*\\
		&(T_{Qi,t}^S-T_{ai,t})+\varrho_{g,t}^{Ri*}*(T_{Qi,t}^R-T_{ai,t}))],  \forall i\in \Psi_{HL}.
	\end{split}
\end{equation}

The electricity energy marginal price of bus $i$ at $t$ is:
\begin{equation}
	\varrho_{p,t}^i=\frac{\partial f^*}{\Delta T_E\partial D_{pi,t}}=\frac{\partial L}{\Delta T_E\partial D_{pi,t}}
	, t\in N_E.
\end{equation}

The payment of load at bus $i$ in the electricity market is:
\begin{equation}
		\pi_{i}^{EL}=\Delta T_E\sum_{t=1}^{N_E}\pi_{i,t}^{PL}=\Delta T_E\sum_{t=1}^{N_E}\varrho_{p,t}^{i*}*D_{pi,t}, \forall i\in \Psi_{EL},
\end{equation}
where $\Psi_{EL}$ is the set of electric load buses.
 
CHP unit $i$ at electric bus $r$ and heat node $k$ receives payment from both the electricity and heat markets:
\begin{equation}\label{paym}
\begin{aligned}
		\pi_{i}^{CHP}&=\pi_{k}^{HS}+\pi_{r}^{ES}=\Delta T_E\sum_{t=1}^{N_E}\varrho_{p,t}^{r*}*G_{pr,t}^*+\Delta T_H\sum_{t=1}^{N_H}\\
		&[(\varrho_{h,t}^{k*}*G_{hk,t}^*)-(\varrho_{g,t}^{Sk*}*(T_{Qk,t}^S-T_{ak,t})+\varrho_{g,t}^{Rk*}*\\
		&(T_{Qk,t}^R-T_{ak,t}))],\forall r\in \Psi_{EB},\forall k\in \Psi_{HN},
		\end{aligned}
\end{equation}
where $\Psi_{EB}$ is the set of electric buses and $\Psi_{HN}$ is the set of heat nodes.
We take CHP units as the general form of generators' payment. Pure electricity/heat-output units, and electric heat pumps are all special examples of (\ref{paym}).
\subsection{Revenue adequacy}
Here we prove that revenue adequacy holds for the two market operators. Using the optimal quantities
solved from (\ref{object}-\ref{cons}), heat energy prices (\ref{he}) and heat grade prices (\ref{hg}), the heat system operator's merchandise surplus at heat period $t$ is:
\begin{subequations}\label{der}
	\begin{align}
		\mathcal{M}_{H,t}&=\sum_{i\in \Psi_{HL}}\pi_{i,t}^{HL}-\sum_{i\in \Psi_{HS}}\pi_{i,t}^{HS}\\
		&=\sum_{i\in \Psi_{HL}}\Delta T_H[(\varrho_{h,t}^{i*}*D_{hi,t})+(\varrho_{g,t}^{Si*}*(T_{Qi,t}^S-T_{ai,t})\\\notag
		&+\varrho_{g,t}^{Ri*}*(T_{Qi,t}^R-T_{ai,t}))]-\sum_{i\in \Psi_{HS}}\Delta T_H[(\varrho_{h,t}^{i*}*G_{hi,t})\\\notag
		&-(\varrho_{g,t}^{Si*}*(T_{Qi,t}^S-T_{ai,t})+\varrho_{g,t}^{Ri*}*(T_{Qi,t}^R-T_{ai,t}))]\\
		&=\sum_{i\in \Psi_{HN}}\lambda_{hi,t}^**(G_{i,t}^*-D_{i,t}^*)\\
		&=\begin{matrix} 
    \underbrace{\boldsymbol{\mu_{t}^{*T}(T_{sa}-T_{a,t})}}\\
    \boldsymbol{CR_{h,t}}
    \end{matrix}
    +\begin{matrix} 
    \underbrace{-\boldsymbol{\lambda_{h,t}^{*T}C_{2}(T_{t-1}^*-T_{a,t})}}\\
    \boldsymbol{IL_{t}}
    \end{matrix}\\\notag
    &+\begin{matrix} 
    \underbrace{\boldsymbol{\lambda_{h,t+1}^{*T}C_{2}(T_{t}^*-T_{a,t})}}\\
    \boldsymbol{IU_{t}}
    \end{matrix}, t\in N_H.
	\end{align}
\end{subequations}
 The heat system operator (HSO)'s merchandise surplus at heat period $t$ can be decomposed by corresponding constraints as congestion rent $\boldsymbol {CR_{h,t}}$, impact from the last/upcoming period $\boldsymbol {IL_{t}}$/$ \boldsymbol{IU_{t}}$. $\boldsymbol{CR_{h,t}}$ and $\boldsymbol{IL_{t}}$ are always non-negative. For detailed derivation of (\ref{der}), please refer to \cite{2015arXiv150203167I}. 

The merchandise surplus of the HSO over $N_H$ heat periods is calculated as:
\begin{equation}
\begin{aligned}
	\mathcal{M}_H&=\sum_{t=1}^{N_H}[\boldsymbol{\mu_{t}^{*T}(T_{sa}-T_{a,t})}]-\boldsymbol{\lambda_{h,1}^{*T}C_{2}(T_{0}-T_{a,1})}\\
	&=\sum_{t=1}^{N_H} \boldsymbol{CR_{h,t}}+\boldsymbol{IL_{0}}\ge 0,
	\end{aligned}
\end{equation}
which is equal to the sum of congestion rent in the heat system over $N_H$ heat periods and the effect of the initial heat state, and they are both non-negative.

Merchandise surplus of electricity market operator at $t$ is:
\begin{equation}
	\mathcal{M}_{E,t}=\boldsymbol{({\sigma_t}^T\overline{L})}\ge 0, t \in N_E,
\end{equation}
which is equal to the congestion rent at $t$ in the EPS.

The merchandise surplus of the electricity market operator over $N_E$ electricity periods is calculated as:
\begin{equation}
	\mathcal{M}_E=\sum_{t=1}^{N_E} \boldsymbol{({\sigma_t}^T\overline{L})}\ge 0,
\end{equation}
which can be guaranteed to be non-negative.

\section{Components of electricity and heat prices}
According to KKT conditions of the optimal dispatch model (\ref{object}-\ref{cons}): $\frac{\partial L}{\partial G_{hi,t}}=0$ and $\frac{\partial L}{\partial G_{pi,t}}=0$. We can decompose the CHP unit $i$'s electricity and heat energy prices into two parts:
\begin{subequations}
\begin{align}
    \varrho_{p,t}^i&=
    \begin{matrix} 
    \underbrace{\eta_{i3}^C+2\eta_{i4}^CG_{pi,t}^*+\eta_{i5}^CG_{hi,r}^* }\\
    MG_E(i,t)
    \end{matrix}
    +
    \begin{matrix} 
    \underbrace{\frac {\boldsymbol{\gamma_{i,t}^TO_{i}}}{\Delta T_E }} \\ 
    CO_E(i,t)
    \end{matrix},\\\notag
    r&= t\frac{N_H}{N_E}, t\in N_E.\\
    \varrho_{h,t}^i&=
    \begin{matrix} 
    \underbrace{\eta_{i1}^C+2\eta_{i2}^CG_{hi,t}^*+\sum_{n=(t-1)\frac{N_E}{N_H}+1}^{n=t\frac{N_E}{N_H}}\eta_{i5}^CG_{pi,n}^* }\frac{\Delta T_E}{\Delta T_H} \\ 
    MG_H(i,t)
    \end{matrix}\\\notag
    +&
    \begin{matrix} 
    \underbrace{\sum_{n=(t-1)\frac{N_E}{N_H}+1}^{n=t\frac{N_E}{N_H}}\frac {\boldsymbol{\gamma_{i,n}^TK_{i}}}{\Delta T_H }} \\ 
    CO_H(i,t).
    \end{matrix}
    ,t\in N_H.
    \end{align}
\end{subequations}
We use $t_E\in N_E$ and $t_H\in N_H$ to distinguish the electricity and heat periods in the following parts.
\subsection{Marginal generation cost}
Marginal electricity generation cost $MG_E(i,t_E)$ define CHP unit $i$'s marginal cost to produce an additional unit of electricity at $t_E$. $MG_H(i,t_H)$ define CHP unit $i$'s marginal cost to produce an additional unit of heat at $t_H$. 

\subsubsection{Co-generation effect} 
CHP unit $i$'s electricity and heat output are coupled by its feasible region. Thus the heat output effects $MG_E(i,t_E)$ through $\eta_{i5}^CG_{hi,t_E\frac{N_H}{N_E}}^*$. And the electricity output effects $MG_H(i,t_H)$ through $\sum_{t_E=(t_H-1)\frac{N_E}{N_H}+1}^{t_E=t_H\frac{N_E}{N_H}}\eta_{i5}^CG_{pi,t_E}^* \frac{\Delta T_E}{\Delta T_H}$.  

\subsubsection{Different time scales} 
Hybrid time scales are adopted in the dispatch of CHP units. CHP unit $i$'s heat output at heat period $t_H$ influences the $MG_P(i,t_H)$ for $t_H$'s corresponding $\frac{N_E}{N_H}$ electricity intervals (from $\frac{N_E}{N_H}(t_H-1)+1$ to $\frac{N_E}{N_H}t_H$). The electricity output of CHP unit for $\frac{N_E}{N_H}$ electricity intervals influences the $MG_H(i,t)$ at $t_H$.

\subsubsection{Different kinds of CHP units}  
Operation states of extraction condensing and back-pressure CHP units are shown in Fig. \ref{fig3}. We can find that for the back-pressure CHP unit, its electricity output also follows the heat dispatch intervals, and $MG_E(i,t)$ keeps the same over $\frac{N_E}{N_H}$ electricity intervals.

\subsection{Energy coupled cost}
Energy coupled electricity and heat cost-$CO_E(i,t)$ and $CO_H(i,t)$ are both related to CHP unit $i$'s feasible region.
When $CO_E(i,t_E)$ is positive, CHP unit $i$ receives more than its marginal cost in electricity market at electricity period $t_E$. And when $CO_H(i,t_H)$ is positive, CHP unit $i$ receives more than its marginal cost in the heat market at heat period $t_H$. 
\subsubsection{Different time scales}
 $CO_H(i,t)$ at $t_H$ is influenced by its electricity output from $t_E=\frac{N_E}{N_H}(t_H-1)+1$ to $t_E=\frac{N_E}{N_H}t_H$. 
\subsubsection{Different kinds of CHP units} 
The two kinds of CHP units have different conditions of  $CO_H(i,t)$ and $CO_E(i,t)$ because of their different feasible regions and possible operation states.

$\bullet$ Extraction condensing CHP unit $i$: When its operation limit (5) is binding, $\gamma_{i,t_E}>0$ at $t_E$. Only when the CHP unit works on the boundary $b$ whose $O_{i,b}\ge 0$, $CO_E(i,t_E)$ is guaranteed to be non-negative. Only when the CHP unit $i$ works on the boundary $b$ whose $K_{i,b}\ge 0$ or inside the feasible region for $\frac{N_E}{N_H}$ electricity intervals, $CO_H(i,t)$ is guaranteed to be non-negative. 

$\bullet$ Back-pressure CHP unit $i$, its inequations constraints (6a-6b) are always both binding since the pair of inequations represent the linear relationship of CHP unit $i$'s electricity and heat power output.  We define $\underline{\gamma_{i1,t}}$ and  $\overline{\gamma_{i1,t}}$ as shadow prices of constraints (6a) and (6b), $\underline{\gamma_{i2,t}}$ and  $\overline{\gamma_{i2,t}}$ as shadow prices of (6c)-(6d):
\begin{subequations}
\begin{align}
    CO_E(i,t)&=\frac{(\underline{\gamma_{i1,t}}-\overline{\gamma_{i1,t}})O_{i,1}}{\Delta T_E}, t\in N_E\\
    CO_H(i,t)&=\frac{\sum_{n=(t-1)\frac{N_E}{N_H}+1}^{n=t\frac{N_E}{N_H}}(\underline{\gamma_{i1,n}}-\overline{\gamma_{i1,n}})K_{i,1}}{\Delta T_H}\\\notag
    &+\frac{\overline{\gamma_{i2,t}}-\underline{\gamma_{i2,t}}}{\Delta T_H}, t\in N_H.
\end{align}
\end{subequations}

 Considering $O_{i,1}>0, K_{i,1}<0$, only when $\underline{\gamma_{i1,t_E}}-\overline{\gamma_{i1,t_E}}>0$ ,$CO_E(i,t_E)>0$. When $\underline{\gamma_{i2,t_H}=0}$ and $\underline{\gamma_{i1,t_E}}-\overline{\gamma_{i1,t_E}}<0$ from $t_E=(t_H-1)\frac{N_E}{N_H}+1$ to $t_E=t_H\frac{N_E}{N_H}$, $CO_H(i,t_H)$ is guaranteed to be non-negative.

\section{Case study}
A CHP system with an eight-bus EPS and a four-node DHS is taken as an example to verify the effectiveness of the proposed pricing rule. There are two CHP units (extraction condensing CHP unit 1 and back-pressure CHP unit 2) and a pure electricity-output generator in the CHP system. We use $\Delta T_H=1h$ and $\Delta T_E=15min$ as the dispatch intervals. Detailed data can be found in\cite{Test}. Its configuration is shown in Fig. \ref{fig 1}. All tests are implemented on a laptop with an M1 central processing unit. The optimization problems are solved by CPLEX solver running on MATLAB.
\subsubsection{Validation of the pricing mechanism}
\begin{figure}[htbp]
	\vspace{-6.0cm}
	\hspace{-0.4cm}
	\includegraphics[width=4.0in]{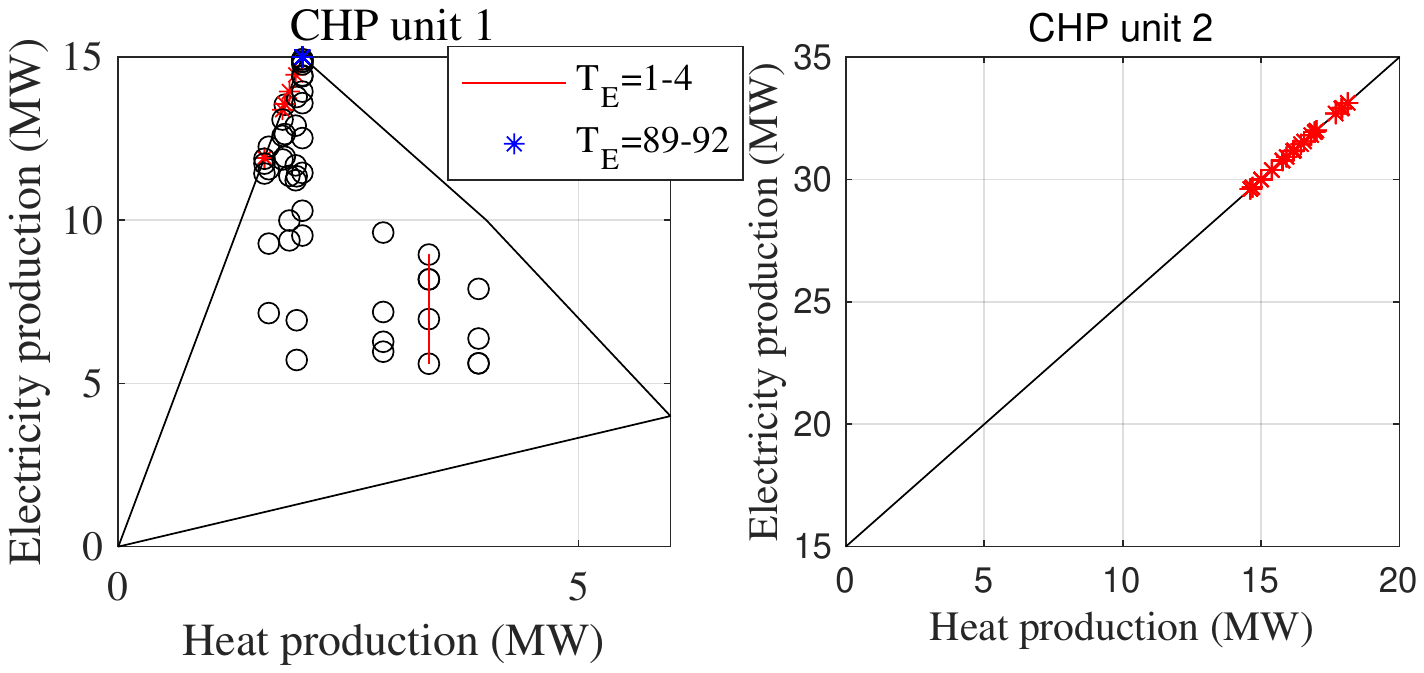}
	\vspace{-6.08cm}
	\caption{Operation states.}
	\label{fig4}
\end{figure}

\begin{figure}[htbp]
	\vspace{-5.2cm}
	\hspace{-0.5cm}
	\includegraphics[width=3.6in]{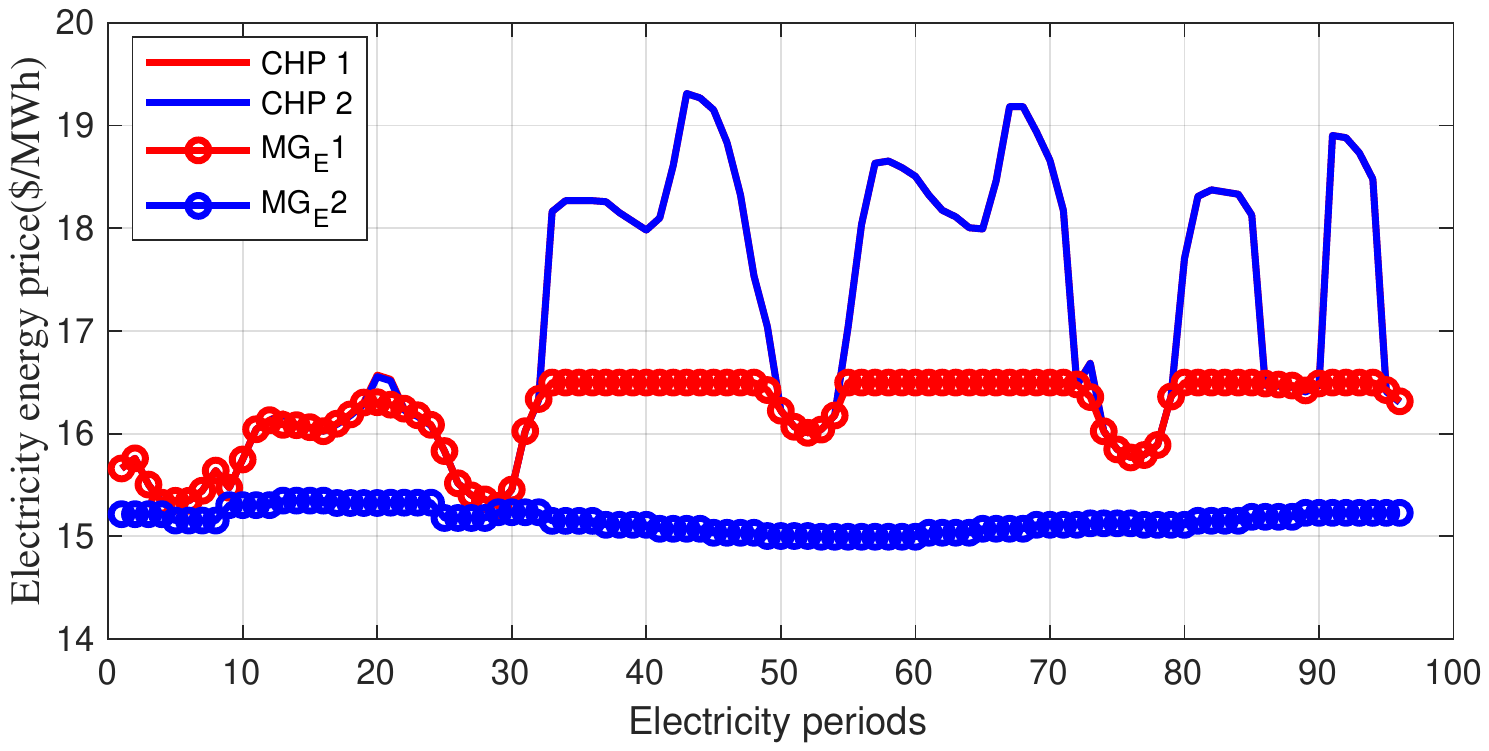}
	\vspace{-5.0cm}
	\caption{Electricity prices.}
	\label{fig5}
\end{figure}

\begin{figure}[htbp]
	\vspace{-5.8cm}
	\hspace{-0.2cm}
	\includegraphics[width=3.6in]{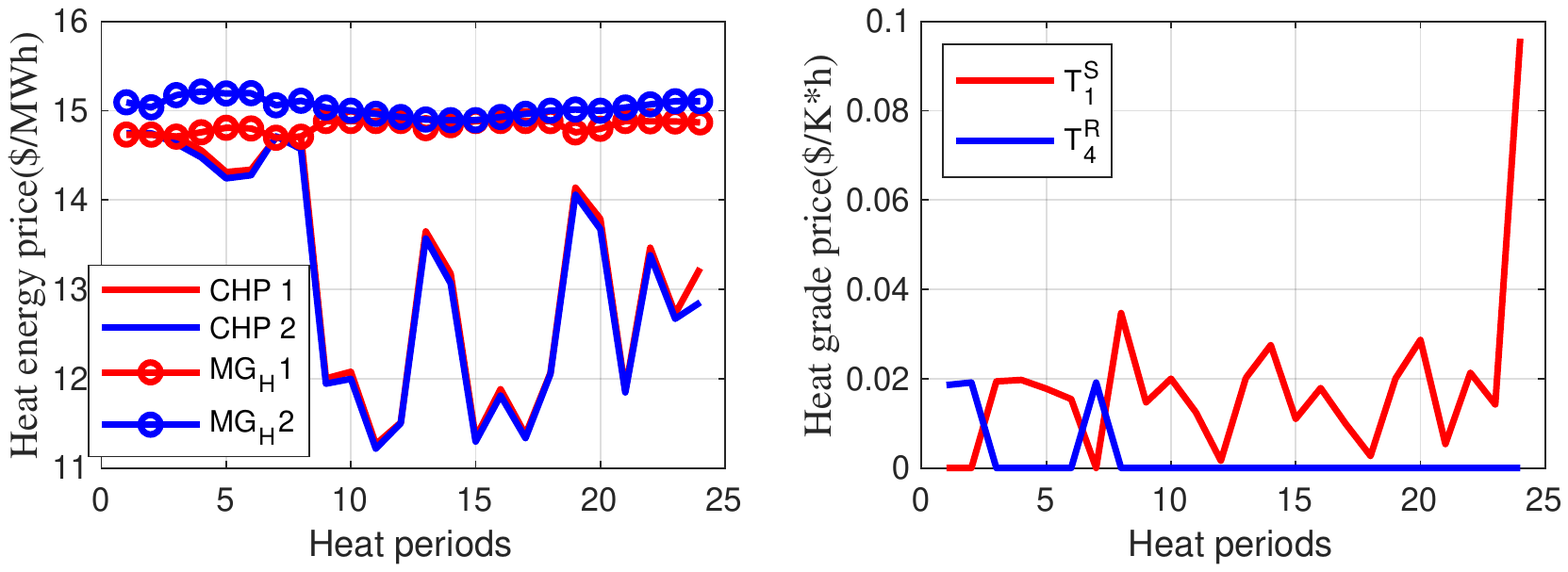}
	\vspace{-5.78cm}
	\caption{Heat prices.}
	\label{fig6}
\end{figure}

The CHP units' operation states are shown in Fig. \ref{fig4}. Their electricity and heat prices are shown in Fig. \ref{fig5} and Fig. \ref{fig6} respectively. 

$\bullet$ Heat grade settlement: From Fig. \ref{fig6}, we can find for some heat periods, heat node 1 and heat node 4 pay for their heat grade requirement because their temperature requirement constraints (8c) are binding. 

$\bullet$ Back-pressure CHP unit 2: Comparing Fig. \ref{fig4} and Fig. \ref{fig5}, we can find that back-pressure CHP unit 2 follows the heat dispatch time scale. But its electricity prices vary across electricity dispatch intervals for its energy coupled cost. 

$\bullet$ Extraction-condensing CHP unit 1: While the extraction-condensing CHP unit 1 has the same heat output and different electricity output over four electricity dispatch intervals. CHP unit 1 always has higher electricity LMP than its marginal electricity production cost while lower heat energy prices than its marginal heat production cost. Because it works inside the feasible region or at the boundary with $O_i>0,K_i<0$. 
\vspace{-0.1cm}

\vspace{-0.7cm}

\subsection{Revenue adequacy}
\begin{figure}[htbp]
	\vspace{-6.0cm}
	\hspace{-1.0cm}
	\includegraphics[width=4.0in]{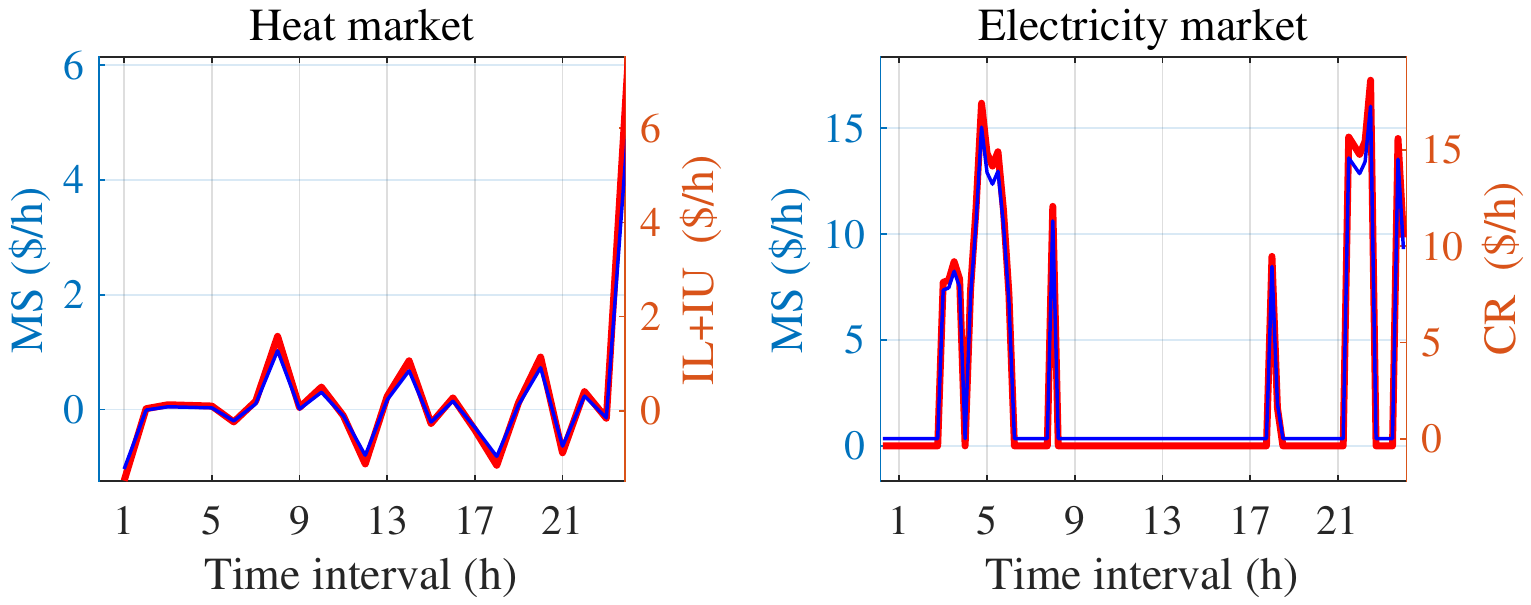}
	\vspace{-6.38cm}
	\caption{Merchandise surplus.}
	\label{fig7}
\end{figure}

Merchandise surplus and its decomposition of the market operators are shown in Fig. \ref{fig7}.
	\vspace{-0.28cm}

Since (8b) is not binding, the heat market operator's merchandise surplus at each heat dispatch period $t$ is equal to the sum of the impact of the last and upcoming period-$IL_t$ and $IU_t$. Thus $\mathcal{M}_{H,t}$ in some heat periods is negative because of the heat transfer’s time-delay effect. The electricity market operator's merchandise surplus at each electricity dispatch period $t$ is equal to the electricity system's congestion rent and is always positive.

The heat market and electricity market operator's merchandise surplus over $N_H$ and $N_E$ periods are \$5.8970 and \$61.6869 respectively, which are both positive. It proves that revenue adequacy holds for both the heat and electricity market operators under the proposed pricing mechanism.

\section{Conclusion}
The proposed pricing mechanism takes heat quality and
different time scales in the CHP system into account. Under
the proposed pricing mechanism, revenue adequacy holds for
both the electricity market operator and heat market opera-
tor. And their merchandise surplus can be decomposed into
interpretable parts. The price linkage between heat and
electricity reflects the energy coupling and different time scales well. Case studies validate the pricing mechanism.

\bibliographystyle{IEEEtran}
\bibliography{IEEEabrv,conference_101719}

\end{document}